# Systems Astrochemistry: A New Doctrine for Experimental Studies

**Nigel J. Mason[1*], Perry A. Hailey[1], Duncan V. Mifsud[1,2], James S. Urquhart[1]**

[1]Centre for Astrophysics and Planetary Science, School of Physical Sciences, University of Kent, Canterbury CT2 7NH, United Kingdom

[2]Institute for Nuclear Research (Atomki), Debrecen H-4026, PO Box 51, Hungary

**\* Correspondence:**
Corresponding Authors
n.j.mason@kent.ac.uk

**Keywords: astrochemistry, interstellar chemistry, molecular astrophysics, systems science, systems astrochemistry, statistical design of experiments**

**Abstract**

Laboratory experiments play a key role in deciphering the chemistry of the interstellar medium (ISM) and the role that product complex organic molecules (COMs) may play in the origins of life. However, to date, most studies in experimental astrochemistry have made use of reductionist approaches to experimental design in which chemical responses to variations in a single parameter are investigated while all other parameters are held constant. Although such work does afford insight into the chemistry of the ISM, it is likely that several important points, such as the relative importance of an experimental parameter in determining the chemical outcome of a reaction and the interaction between parameters, remain ambiguous. In light of this, we propose adopting a new 'systems astrochemistry' framework for experimental studies which draws on current work performed in the field of prebiotic chemistry, and present the basic tenants of such an approach in this article. This systems approach would focus on the emergent properties of the chemical system by performing the simultaneous variation of multiple experimental parameters and would allow for the effect of each parameter, as well as their interactions, to be quantified. We anticipate that the application of systems science to laboratory astrochemistry, coupled with developments in hyphenated analytical techniques and data analytics, will uncover significant new data hitherto unknown, and will aid in better linking laboratory experiments to observations and modelling work.

## 1    Introduction

### 1.1    The Science of Astrochemistry

One of the unexpected discoveries arising from the development of radio astronomy in the 1930s was the discovery of molecules in the interstellar medium (ISM), since up to then all spectroscopic signatures had been atomic in nature. The clearly identified spectroscopic signatures of the diatomic radicals CN and CH (McKellar 1940) and OH (Weinreb et al. 1963) clearly indicated that larger parent molecules were present. This was subsequently confirmed by the discovery of formaldehyde by Snyder et al. (1969) which itself suggested that organic, polyatomic molecules may be present in the ISM. The presence of a particular group of carbon-containing molecules – the polycyclic aromatic hydrocarbons (PAHs) – was soon after proposed to explain astronomical observations of high photo-absorption in the ultraviolet region; the so-called 'UV extinction bump' (Stecher 1965, 1969). With an increasing number of radio telescopes and greater sensitivities becoming accessible, the number of molecules identified in the ISM has steadily grown to over 200,[1] with various structures having been identified including the fullerenes $C_{60}$ and $C_{70}$ (Cami et al. 2010), the simplest sugar glycolaldehyde (Jørgensen et al. 2012), and chiral molecules such as propylene oxide (McGuire et al. 2016).

The existence of organic molecules in turn provided the rationale for a new hypothesis on the origins of life on Earth in which the necessary chemical ingredients were delivered to an early Earth by extra-terrestrial bodies such as comets, asteroids, and meteorites thus seeding the Earth with prebiotic molecules. These prebiotic molecules were subsequently assembled into more complex structures on Earth, contrasting with the Miller-Urey hypothesis of such molecules having been created on Earth by, for example, atmospheric chemistry (Miller 1953, 1955). The discovery of a rich molecular inventory, which included amino acids, in many meteorites supported this extra-terrestrial origin of material and thus suggested that the chemical ingredients of life may indeed be delivered to any planet as a consequence of Solar System formation (Burton et al. 2012 and references therein). The development of life is then dependent on the ability of the planet to assemble this prebiotic molecular material into more complex structures, and not on its ability to sustain a *de novo* synthesis of the chemical ingredients of life.

These discoveries required an explanation of the chemistry that could lead to the formation of molecules in what were, until then, regions of space regarded to be both too cold and too 'empty' to sustain chemistry. The study of chemistry in such regions was soon entitled 'astrochemistry' and was simply defined as the chemistry prevalent in astrophysical settings. With the emergence of new studies and space missions, however, the term has come to refer to various different fields within the space sciences. For example, it is often used to embrace the study of chemistry on other planetary bodies and their moons as well as on comets and, more recently, the predicted chemistry on exoplanets. It has also embraced and often absorbed 'cosmochemistry' or 'chemical cosmology' which, if strictly following the Greek, is the study of the chemical composition of matter in the Universe and the processes that led to those compositions, and is still a term widely used by the meteoritic and isotope geochemistry communities.

In this review, we will define the term 'astrochemistry' to be the study of molecules in the ISM both in terms of their detection and the explanation of the physico-chemical processes that lead to their formation. We will outline the current state of knowledge and discuss how laboratory studies are being used to explain observations, test hypotheses on the routes of molecular synthesis, and validate models of star and planet formation which incorporate ISM chemistry. However, many of our comments and conclusions are equally valid for studies of any extra-terrestrial icy body such as comets, Kuiper Belt Objects (KBOs), and the Jovian and Saturnian satellites. We wish to introduce a new doctrine in the design and conduct of laboratory studies that draws upon the experience of other research fields and exploits the methodology of systems chemistry to address what is a complex, multi-parameter problem that has led to a plethora of laboratory studies that often report conflicting results.

---

[1] For a regularly updated list of molecules detected in the ISM, refer to the Cologne Database for Molecular Spectroscopy (https://cdms.astro.uni-koeln.de/classic/molecules).

We thus wish to address a major problem facing the astrochemistry community as it expands and more experimental groups become active; namely the mantra that 'less is more' in data generation. Experimental studies can not be cross-correlated and, to date, none provides a truly representative analogue of the conditions in the ISM. We propose some new protocols and methodologies that we believe, if adopted by the community, will lead to a more coherent set of data. Such data can be cross-correlated across the community and will more clearly define the parameter space, not only for conducting experiments, but also for testing hypotheses and models as well as formulating predictions that can be validated by observations made by the new generation of telescopes, including the Atacama Large Millimeter/Submillimter Array (ALMA) and the James Webb Space Telescope (JWST).

## 1.2 Current Understanding of Molecular Synthesis in the ISM

The major challenge of astrochemistry is to explain the formation of such a rich interstellar molecular inventory extending from the large PAHs and fullerenes to the simplest (and most common) molecule which is $H_2$. The conditions for molecular formation, however, are such as to appear to preclude chemistry at low particle number densities (due to infrequent collisions precluding three-body gaseous reactions) and extremely cold temperatures (due to reactants having insufficient energy to overcome reaction barriers).

Recognizing that the most common state of matter in the Universe is plasma, the possibility of ion-molecule reactions as a possible route to molecular formation has been studied since the 1970s. Using ion storage rings and adapted molecular beams, many ion-molecule reactions have been studied and have been shown to be both barrierless and have kinetic rate constants which increase at lower temperatures; such features are commensurate with the requirements for chemistry in the ISM (Larsson et al. 2012, Geppert and Larsson 2013). Observations of molecular ions such as $H_3^+$, $HCO^+$, $HOC^+$, and $C_3H^+$ (Larsson et al. 2008, Gerin et al. 2019) further confirm the presence of such chemistry.

However, gas-phase chemistry such as this can not account for the formation of the ubiquitous $H_2$ nor the abundances of any of the observed polyatomic species. The solution to this problem is the inclusion of solid-state chemistry: although much of the Universe appears devoid of stars, in reality they are only hidden by large 'dust clouds' along the line of sight. Interstellar dust is formed as a product of stellar death and is typically comprised of micron-sized particles of carbon or silicates. If the dust cloud is thick enough, any light from stars behind it will be completely blocked leading to the appearance of dark and empty areas. It is from these dust clouds that, through gravitational collapse, stars and planetary systems form.

Importantly, it is also on the surfaces of the dust grains that molecular formation may take place. For example, such grains provide the surface upon which two hydrogen atoms in the ISM can meet and combine. Such a reaction has been tested in the laboratory and has been shown to not only produce $H_2$ but to also produce it in the ro-vibrational states observed in the dust regions of the ISM (Wakelam et al. 2017). The condensation of simple molecules formed in the ISM (e.g., by ion-molecule reactions) on such dust leads to the formation of an ice mantle (Fig. 1) within which chemical evolution may take place.

Within its lifetime (which is typically on the scale of $10^5$-$10^6$ years), any ice-covered dust grain may experience various methods of processing which lead to the synthesis of more complex material. Such processing includes: (i) atom (or radical) addition chemistry from incident species impacting on the grain and its previously accreted layers, (ii) thermal processing of the ice, for example during gravitational collapse and stellar evolution, (iii) photochemistry induced by the interstellar radiation field (ISRF), radiation from nascent stars, or Lyman-α photons emitted as a result of cosmic ray interaction with $H_2$, (iv) radiation chemistry induced by galactic cosmic rays and X-rays, and (v) shock processing induced by supernovae. These processing types are summarized in Fig. 2.

## 2 Laboratory Studies of Ice-Grain Chemistry: State of the Art and the Challenges of Solid-Phase Laboratory Astrochemistry

Laboratory studies of molecular synthesis in ices have been prevalent since the mid-1970s when Mayo Greenberg and co-workers at Leiden in the Netherlands explored such synthesis in ice films deposited on a variety of substrates using Lyman-$\alpha$ radiation generated from hydrogen discharge lamps. It was immediately clear that UV irradiation of even simple mixtures of molecules can produce a large number of products. In a pioneering paper, Muñoz-Caro et al. (2002) showed that Lyman-$\alpha$ irradiation of an ice mixture containing the simplest molecules believed to be present in dust grain ice mantles (i.e., $H_2O$, $CO$, $CO_2$, $NH_3$, and $CH_3OH$) led to the synthesis of 16 amino acids some of which were chiral in nature.

Such experiments suggest that the formation of amino acids in the ISM is possible and thus explains their presence within meteorites, supporting the idea that prebiotic molecules could have been delivered to the early Earth by meteorites, cometary dust, or interplanetary dust particles during the Late Heavy Bombardment approximately four billion years ago. Further studies on the residues left after thermal desorption of the volatile ices revealed that the product molecules are amphiphilic and are typical of molecules that make up cell membranes, being themselves capable of forming membrane-like vesicles when immersed in water. Vesicle formation is thought to be critical to the origin of life as these structures provide an isolated environment in which life processes are protected from the surrounding medium (Deamer et al. 2002).

Such pioneering investigations have led to a plethora of experiments and active research programs in many countries. The production of complex molecules in irradiated ices has now been demonstrated for many years and has been extended to many different irradiation types, including X-rays, passive ion beams (e.g., helium ion beams), and reactive ion beams (e.g., carbon ion and proton beams). These primary radiation sources have all been shown to produce a cascade of secondary electrons with a wide range of energies ranging from that of the primary source down to meV. Indeed, many thousands of secondary electrons may be produced along the track of a single ion. Electron irradiation of astrochemical ices has revealed that the synthesis of complex molecules is possible even at very low energies, including those energies well below the ionization potential of the target molecule or the lowest excited state via the process of dissociative electron attachment (Mason et al. 2014).

Today, the literature contains many hundreds of publications reporting results from the processing of ices composed of different combinations of molecules irradiated by many different sources. Photochemistry has been explored using the wide range of wavelengths provided by synchrotron radiation (e.g., de Marcellus et al. 2011, Lo et al. 2014) and ion irradiation chemistry has been performed using multiply charged ions including $Zn^{26+}$ and $Xe^{23+}$ with energies from a few keV to a few hundred MeV (e.g., Moore et al. 1983, Ding et al. 2013, Mejía et al. 2013). Atom addition experiments have revealed that many of the simplest molecules such as $H_2O$, $NH_3$, $H_2S$, and $CH_4$ are formed by successive hydrogen atom additions to accreted oxygen, nitrogen, sulfur, and carbon atoms (Linnartz et al. 2015). Formaldehyde is, in turn, formed by hydrogen addition to adsorbed CO and can itself be further hydrogenated to yield methanol (Chuang et al. 2016).

However, although such experiments clearly demonstrate the synthesis of molecules and provide routes by which they can form, they are often contradictory both in terms of the product molecules formed and their concentrations. No two experiments replicate exactly the results of the other, even if they appear to have been performed under similar conditions. Additionally, some of these studies report results obtained under experimental conditions not necessarily representative of the ISM. Therein lies a major challenge of laboratory astrochemistry: the replication of results and of interstellar conditions. This lack of replication may be ascribed to the many variables and parameters in a solid-state astrochemistry experiment meaning that not only do the results in different laboratories vary, but differences are observed when repeating experiments in a single laboratory due to systematic factors in experimental apparatus and procedures.

For instance, the morphology of an ice is expected to play a major role in an astrochemical reaction, since it determines the position and mobility of reactants (Hornekær et al. 2003, 2005). Ice morphology is dependent on the temperature of the substrate upon which molecules are deposited, with different phases (crystalline and amorphous) being preferentially formed at different temperatures. Another parameter which may influence the morphology of the ice is the time taken for its deposition. Experimental evidence has suggested that ices deposited over longer time-scales may have a higher degree of crystallinity compared to those which are deposited rapidly (Holtom et al. 2006, Mason et al. 2006).

The morphology of the ice is also somewhat dependent upon the structure of the substrate. In the laboratory, ices are typically deposited onto flat substrates made out of a material which is transparent to or reflective of the particular wavelength of light used for analytical spectroscopic monitoring. For example, IR-transparent materials such as zinc selenide or cesium iodide are commonly used, as are reflective gold surfaces. Such materials do not simulate interstellar dust grains well, as these actually possess a highly irregular morphology and, as previously mentioned, are largely made of amorphous magnesium and iron silicates or carbon-based structures (van Dishoeck 2014).

Indeed, it is this highly irregular morphology in which dust particles present structural features such as steps, pores, and terraces which is thought to be the main reason why interstellar dust grains are such productive chemical factories (Dulieu et al. 2013). Although previous studies have alluded to the importance of the role of dust grain morphology in astrochemical reactions, there is a scarcity of studies which have considered this experimentally. One study by Mason et al. (2008) showed that the IR spectrum of hexagonal crystalline $H_2O$ ice deposited over soot particles suspended in an ultrasonic trap differed somewhat to that of the same ice deposited over flat fluoride substrates traditionally used in laboratory astrochemistry. This is significant, as it is believed that soot particles produced during combustion processes are morphologically similar to the carbonaceous dust grains encountered in the ISM (Cataldo and Pontier-Johnson 2002).

A major experimental parameter which is often under-reported in the literature is that of pressure; specifically, the background pressure in experimental chambers used to carry out astrochemical studies. Ideally, this pressure should be below $10^{-9}$ torr so as to ensure that the residual gas composition is mainly due to $H_2$ after the system is baked out. However, several experiments (particularly those making use of central facilities in which access to radiation sources is limited) are performed at base pressures greater than $10^{-9}$ torr thus allowing residual gases to be frozen out by the cryostat resulting in their possible integration into the ice mantle. The minimum achievable base pressure of an experimental set-up is dependent upon the design and cost of the vacuum system and the number of additional ion pumps used. However, although certain experimental set-ups are capable of achieving lower background base pressures (e.g., $10^{-10}$ torr), this is still orders of magnitude higher than what is expected in the ISM.

Another parameter which may influence the outcome of an astrochemical reaction is that of temperature. Experiments conducted at a set temperature relevant to interstellar molecular clouds (e.g., 20 K) may be indicative of one chemical outcome within icy grain mantles. However, it is known that radical species are significantly more mobile throughout the ice matrix at such a temperature compared to lower ones (Garrod and Herbst 2006, Watanabe et al. 2010). For example, at temperatures below 20 K, $O_2$ may dimerize to form an $O_2$-$O_2$ complex which is easily dissociated leading to efficient $O_3$ formation. At temperatures above 30 K, however, ozone formation becomes less efficient as the dimer complex is no longer formed and oxygen atoms have to diffuse through the ice before reacting (Sivaraman et al. 2007). At even higher temperatures, thermal reactions within the ice matrix become more favourable and may thus begin to offer new, competing reaction pathways (Loeffler and Hudson 2010, 2016).

The thickness of the ice used in an experimental study is also of consequence. Oftentimes, thicker ices are used so that the penetrating radiation energy is absorbed by the ice itself thus ensuring that the deposition substrate plays no role in the observed chemistry. Such thick ices are good analogues of planetary and lunar ices, but ice grain mantles in the ISM are significantly thinner and may not even fully cover the grain. As such, chemistry in the ISM may involve reactions resulting from direct interaction between the incident radiation and the dust grain. Chemistry arising from ion irradiation is particularly sensitive to ice thickness since the maximum energy absorption occurs at the so-called 'Bragg peak'. If the Bragg peak is located in the substrate or at the substrate-ice interface then new chemistry may occur as a result of electrons from the substrate entering the ice mantle.

In seeking to mimic the conditions present in the ISM the one parameter which is impossible to replicate is that of time. Compared to experiments performed in the laboratory (which typically occur over a period of a few hours), the processing of ice grain mantles in space occurs on an almost infinite time-scale, with their formation and processing taking place over $10^5$-$10^6$ years. Atom or molecule deposition occurs entity by entity with each atom or molecule having time to diffuse across the surface of the grain or into the ice matrix before another atom or molecule deposits. On the other hand, laboratory ice depositions are typically conducted in such a way that each atom or molecule is likely to collide with another *en route* to deposition. Reference has already been made to the fact that the rate of deposition may influence the structure and morphology of the resultant ice, with longer deposition times having been shown to increase the degree of crystallinity (Holtom et al. 2006, Mason et al. 2006). Therefore, given the astronomical time-scales encountered in the ISM, it is probable that most ice grain mantles possess some degree of crystallinity or, at the very least, polycrystallinity. The large time intervals between events in the ISM may also lead to the stabilization of radical species in the ice and the dissipation of the energy of excited species; something not replicated by laboratory studies.

However, it is important to note that such time delays imply that space chemistry likely occurs sequentially on an event-by-event basis allowing for the effects of each process to be added separately. Thus, extensive knowledge may be gleaned from studies looking into sequential processing of different types (e.g., ultraviolet photon irradiation followed by ion irradiation). Furthermore, it is possible that galactic cosmic ray irradiation of interstellar ice grain mantles is accompanied by some Lyman-$\alpha$ photolysis, thus providing a motivation for simultaneous processing of different types (e.g., simultaneous ultraviolet photon and ion irradiation). However, although several systematic studies comparing the physico-chemical effects of different irradiation types are available in the literature (Yakshinsky and Madey 2001, Baratta et al. 2002, Muñoz-Caro et al. 2014, Arumainayagam et al. 2019), to the best of our knowledge experimental studies reporting the results of sequential or simultaneous processing by different methods are very rare.

The duration of laboratory experiments is, in general, rather short and rarely exceeds a few hours irrespective of the final fluence of photons or charged projectiles delivered. Thus, many experiments reveal only the formation of 'first-generation' molecular species produced by re-combination of fragments derived from the dissociation of the molecular species initially deposited onto the substrate. Typical product yield dependencies on fluence are depicted in Fig. 3 which shows that, after an initial growth the curve plateaus. However, longer-term irradiation may lead to the first-generation product species being themselves dissociated resulting in second-generation products. This is especially interesting if the ice is heated thus driving off volatile species and leaving behind a refractory mantle of complex organic molecules (COMs) which, once further irradiated, may be either broken down into simpler molecules or react with other species to form even more complex structures. Such a processing methodology is representative of an ice grain mantle being irradiated in the ISM and then being further processed as the original dust cloud collapses to form stars and planetary systems during which the ice is sublimated to leave behind grains with residues rich in COMs.

Finally, it should be stressed that when comparing laboratory data with observations, the latter are themselves an ensemble of many different grains processed by many different methods forming an averaged measurement of products, whereas the former represent a discreet snapshot of one grain mimic having undergone a discrete and specific processing type. In order to provide a direct comparison with observations one should perhaps provide a statistical ensemble of different experiments to reflect the different processing types and ice mixtures in a defined region of the ISM. Ideally, the formation of certain molecules may evidence distinct reaction pathways and conditions. For instance, the processing of $CO_2$:$NH_3$ ices yields both carbamic acid $H_2NCOOH$ and ammonium carbamate $NH_4(H_2NCO_2)$; however, which molecule is the major product is dependent upon whether it was $CO_2$ or $NH_3$ which was in excess in the original ice mixture (James et al. 2020). In a similar vein, the processing of polar ices may result in different chemical outcomes compared to the processing of apolar ones, and perhaps some products will be shown to be a marker of ultraviolet photon processing while others may indicate ion irradiation.

Hence, by understanding the route of formation of each interstellar molecular product and quantifying the relative importance of each parameter involved in its synthesis, it may be possible to extensively characterize the physico-chemical properties of different regions in the ISM, as well as different phases of stellar and planetary system evolution from dust clouds. However, in order to achieve such an exhaustive characterization a new approach to experimental design and conduct is needed which carefully analyzes the dependence of the synthesis of any molecular product on each parameter of the experiment. This multi-parameter approach requires a formal statistical design of the astrochemical experiment and should adopt a 'systems chemistry' approach. In systems chemistry, the focus is not on individual chemical components but rather on the overall network of interacting molecules and their emergent properties (Ludlow and Otto 2008). In the rest of this article, we will discuss the concepts of systems chemistry before elucidating how we may develop a 'systems astrochemistry' framework to aid in our understanding of chemistry in the ISM and the formation of COMs with prebiotic significance.

# 3    Systems Chemistry

A complex system is defined as a collection of interdependent components capable of interacting with each other which is difficult to model due to the number and magnitude of competing interactions and relationships (Siegenfeld and Bar-Yam 2020). The study of complex systems typically makes use of a holistic, systems-wide approach in which the focus is placed on the emergent properties of the system as a whole, rather than its constituent parts and their simple interactions (Whitesides and Ismagilov 1999, Ross and Arkin 2009). Complex systems analysis has proven attractive to a variety of academic fields and has been adopted by many of them; the most relevant to this article is that of chemistry for which a 'systems chemistry' paradigm has emerged (Ludlow and Otto 2008).

Systems chemistry frameworks seek to consider multiple variables and parameters simultaneously and focus on the emergent chemical products deriving from the complex system under investigation (Ludlow and Otto 2008, Li et al. 2013, Mattia and Otto 2015). This is in contrast to the reductionist approaches which have traditionally sought to understand bond formation from a simple, linear perspective, and which became widespread due to limitations in available analytical equipment and methodologies, specific requirements related to reaction yields and product purities, and the belief that understanding smaller and simpler components of a system may provide some insight into more complex chemistry.

Although still in its infancy, a systems chemistry framework has already been adopted in the field of prebiotic chemistry to study the assembly of chemical sub-systems into an overall larger system with a focus on its emergent physico-chemical properties (Powner and Sutherland 2011). This has led to a fundamental change in our approach to understanding the chemistry of life's origins by considering molecules beyond those that are used by extant biology including prebiotically relevant molecules that existed alongside biogenic ones (Krishnamurthy 2020). Results from such prebiotic systems chemistry work have been far-reaching, with the emergence of RNA being demonstrated to not be a simple outcome from reactions between plausible prebiotic precursors but rather a systems chemistry emergence from a library of molecules (Kim et al. 2017). Additionally, problems related to prebiotic ribonucleotide selection have been shown to be circumvented by a system-wide sequestration of glyceraldehyde by 2-aminothiazole (Islam and Powner 2017, Islam et al. 2017).

If adapted to suit laboratory research in astrochemistry, a systems chemistry approach would allow for the effects of multiple experimental parameters and variables to be considered and quantified, which may yield significant insights into the formation of COMs. As discussed above, laboratory astrochemistry as it has been practiced thus far has largely adopted an approach in which investigations center either on simple ices irradiated by a single processing type, or on the formation of complex materials via the irradiation of mixed ices and substrates.[2] The challenge thus is to learn from such a working paradigm to develop a coherent understanding of the formation of COMs in a structured yet relevant manner. Indeed, the need for a more systematic approach has already been referred to in a number of publications (e.g., Carota et al. 2015, James et al. 2020, Gentili 2020).

We therefore propose that a systems astrochemistry framework be adopted in laboratory astrochemistry experiments, wherein multiple variables and parameters are studied simultaneously under conditions relevant to the ISM. That is to say, the physico-chemical constraints, role of ice morphology and polarity, grain catalysis, multiple processing methods, and the recycling of molecular material during cloud evolution should all be simultaneously studied in a structured systematic design to produce the conditions under which the chemical inheritance of COM formation is explored. Adopting such a systems astrochemistry approach by re-purposing practices from the field of prebiotic systems chemistry and focusing on the emergent properties of the system as a whole could reveal insights into the formation of COMs not gleaned from a linear and sequential approach. The remainder of this section of the article is devoted to discussing the themes relevant to a proposed systems astrochemistry experiment.

---

[2] Several excellent reviews are available on the current state of the art of laboratory astrochemistry (e.g., Öberg 2016, Arumainayagam et al. 2019).

## 3.1 Astrochemical Design Space

When adopting a systems astrochemistry approach to a laboratory investigation, the concept of the astrochemical design space must be considered. This space should look to mimic conditions in the ISM, considering all relevant factors such as the temperature(s) under investigation, the processing type (e.g., photon or ion irradiation), the substrate (including its composition and morphology), the polarity, morphology, and chemical composition of the ice analogue, the time for various factors including deposition and processing, the type of chemistry occurring (e.g., photochemistry, electron-induced chemistry, etc.), and any other relevant physical and chemical constraints. As such, the experiment should be considered as an input-process-output (IPO) sequence (Fig. 4) in which the emergent properties (output) are the result of controlled processing of molecules and substrates (input) under conditions relevant to the ISM (process).

## 3.2 Formal Experimental Design: Interstellar Conditions Replication and Parameterization

Adopting a systems astrochemistry approach requires the consideration of all relevant factors listed in the design space so as to allow the interactions between said factors to be investigated. Laboratory astrochemistry experiments have typically employed a one-factor-at-a-time (OFAT) framework, in which one factor is varied whilst all others remain constant. For instance, studies investigating the temperature dependence of product formation during a given ion irradiation of a molecular ice with a pre-defined composition operate under an OFAT framework. Such experiments are useful as they are simple to execute and comparatively quick to perform, which is important when access to experimental facilities is limited. However, a major disadvantage of studies making use of an OFAT approach is that the relationships between parameters can not be investigated. Furthermore, OFAT studies lack precision in comparison to more formal experimental designs with an equal number of experimental runs and can miss the optimal setting of a factor for a particular output (e.g., product formation).

Instead, we propose that laboratory astrochemistry studies should embrace formal experimental designs as part of a systems astrochemistry framework. Designs such as screenings or response surface designs, for instance, allow all relevant factors to be built into the experiment and varied in a series of experimental runs, after which the outcome from each run is fitted against the analytical and product responses. At face value, formal experimental designs may require more experiments to be performed than an OFAT approach but come with the benefit of revealing the interactions between variable factors and their impact on the emergent properties of the system. In contrast, an OFAT approach would not yield information on such interactions and may result in a potential misinterpretation of the emergent properties. This is highlighted by the fact that there exist formal experimental designs, such as the Plackett-Burman design, in which the number of experimental runs is similar to that when using an OFAT framework but more information on the effect of each individual factor is revealed.

Simple screening designs such as full or partial factorial designs can be employed under a systems astrochemistry approach, although care must be taken to manage the number of experimental runs required based on the number of factors and levels of investigation. Fig. 5 represents a simple full factorial experimental design for three investigated factors at two levels for a total of eight experimental runs; each represented by a corner of the experimental design space. As an example, Fig. 5 may represent a systems astrochemistry ion beam irradiation study in which the three investigated factors are temperature, ion beam energy, and ice thickness. Each of these factors may be set at one of two levels: a 'low' level (e.g., low temperature, low ion beam energy, and thin ice) or a 'high' level (e.g., higher temperature, higher ion beam energy, and thicker ice). A fuller description of such a study is given in Section 4.1.

The number of experimental runs required for any full factorial design is easily calculated as the number of levels investigated raised to the power of the factors included in the design space. As experimentation becomes more complex and more variables and levels are introduced, fractional factorial designs may be employed to constrain the number of experimental runs to manageable levels. Many other experimental designs which may be invoked when making use of a systems astrochemistry approach are also available, however a complete description of these would go beyond the scope of this article. Instead, we direct the interested reader to the work of Deming and Morgan (1993) for a comprehensive overview.

### 3.3 Chemical Inheritance and Recycling

As the dense dust cloud in which interstellar ice grain mantles find themselves evolves via gravitational collapse, a nascent star is formed which may potentially develop its own planetary system before returning copious molecular material to the cloud upon its death. During this cosmic cycle, ice grain mantles are subjected to changing physico-chemical conditions and processing methods. In order to fully assess the formation of COMs in terms of abundance and location or timing of formation, a systems astrochemistry experiment should build all such variables into the design space.

Presently, the vast majority of laboratory astrochemistry experiments make use of a single method of ice processing. As such, only the chemical outcomes of very specific circumstances are deciphered. However, in order to obtain a more realistic view of interstellar chemistry (refer to our discussion in Section 2), sequential irradiation of different processing types, co-irradiation of different processing types, and cycles of heating, deposition, and cooling need also to be built into the experiment design. Understanding the influence of such changing parameters is crucial, as it is already known that different processing methods result in different chemistry, and may thus produce a different chemical feedstock for subsequent reactions (Mullikin et al. 2018).

As such, the production of COMs in the ISM is not a linear or orthogonal process and is likely to involve several stages of production and destruction as the ice grain mantles experience different interstellar conditions (Fig. 6). Investigations of multi-component systems with multiple processing types in cycled experiments will produce further information on astrochemical sub-systems which, when combined together, will reveal the interplay between various molecules yielding COMs of prebiotic relevance.

### 3.4 Astrochemical Pathways and Combinatorial Approaches

Given that astrochemical reactions involve both simple and complex molecules, and that COM formation is not a linear or orthogonal process, the concept of chemical formation in the ISM resulting from component chemistries should be explored so as to further develop astrochemical pathways. This inventory of molecules provides a pseudo-combinatorial astrochemistry approach which may be used to develop our understanding of competing and complementary astrochemical pathways during molecule formation in the ISM. To illustrate this point, the simple energetic processing of single component or two-component ices generates a wealth of small molecules (e.g., Henderson and Gudipati 2015), as shown in Fig. 7. This pool of molecules is then capable of undergoing further energetic processing to produce different, and possibly larger, molecules of interest. This example demonstrates the vast array of molecules which could be formed from relatively simple ices. As such, applying combinatorial strategies for the selection of libraries of molecules in a systematic manner could reveal interesting chemistries which could be used as chemical markers of different ISM environments. A similar approach has been demonstrated recently in a prebiotic chemistry experiment wherein a fully automated system explored complex mixtures over extended timeframes, all the while taking a systems view of the multi-component chemistry (Asche et al. 2021).

## 4  Adopting Systems Astrochemistry in the Laboratory

### 4.1  Parameterization and Statistical Design: A Systems Astrochemistry Thought Experiment

Reference has already been made to the OFAT approach traditionally used in most laboratory studies, as well as the comparative advantages offered by formal statistical experimental designs which study the effect of varying multiple experimental parameters simultaneously (see our discussion in Section 3.2). Briefly, the principal advantages of formal statistical designs include a more precise estimation of the effects of each factor and parameter, a fuller mapping of the experimental design space, and the possibility of analyzing the interactions between factors.

To illustrate the use of a statistical experimental design under a systems astrochemistry approach and its advantages, in this section we will describe a thought experiment in which an interstellar ice analogue is processed using such an approach and then compare that with a similar ice processed under an OFAT approach. Consider first a simple, single component (i.e., pure) ice which is processed by a single method (e.g., ultraviolet photons); such a scenario is fairly common within laboratory astrochemistry. Let us now define five parameters which may influence the outcome of the system, namely the ice thickness, the morphology of the ice, the energy of the processing method, the final fluence reached during processing, and the temperature of the system. Finally, let us consider two levels for each of these parameters (low and high levels). Table 1 provides an example of these parameters and levels.

A full factorial design would thus require a study involving $2^5 = 32$ experimental runs. Each of these experimental runs would consider a specific combination of parameters and all possible combinations would be investigated across the 32 runs (Table 2). It is also common to add on a number of center point analyses to this experimental run, so as to better identify the influence of each parameter on changing from one level to another. Thus, for example, three center point analysis runs could be performed for 1.0 μm-thick ices processed at 60 K by a 50 keV energetic processing method to a final fluence of $10^{15}$ cm$^{-2}$. This would take the total number of experimental runs to 35.

If the same ice processing were carried out under an OFAT approach, however, the number of experiments would be reduced to 10 as each level for a given parameter would be investigated only as all other parameters were held constant. It is thus evident that an OFAT approach, although providing some insight into the chemical reactivity induced by ice processing, can not assess the effect of various parameter combinations on the emergent properties of the system. This is best visualized through an astrochemical design space. For ease of visualization, we depict the experimental runs involved in a full factorial experimental design in which three parameters are investigated at two levels on an astrochemical design space and compare this with a design space for a three-factor OFAT study (Fig. 8).

Upon completion of the experimental runs in a statistical experimental design such as the one described above, the documented responses may be analyzed in a variety of methods, including the use of multi-linear or polynomial fits, Pareto plots, multi-dimensional scatter plots, or some other technique thus allowing for the influence of each parameter and each combination of parameters on the emergent properties of the system (e.g., abundance of an expected COM) to be determined and quantified. An example of this is given in Fig. 9.

## 4.2    Analytical and Processing Considerations

Although reference has already been made to the standard analytical techniques used in laboratory astrochemistry experiments, it is perhaps worthwhile to spend some time discussing their roles in systems astrochemistry. The most commonly used *in situ* analytical techniques are quadrupole mass spectrometry (QMS) and Fourier-transform infrared spectroscopy (FTIR) operating either in transmission or reflectance mode. Other techniques are also becoming more popular: the use of millimeter/submillimeter and terahertz spectroscopy, for instance, will likely help bridge results from the laboratory to observational studies (Yocum et al. 2019, Widicus Weaver 2019).

Although *in situ* techniques are useful in identifying functional groups present in the ice and sputtered or desorbed molecules in the gas-phase, it is possible that some products form a refractory solid residue which is too complex for its individual components to be resolved via infrared spectroscopy. As such, the chemistry resulting in the formation and depletion of this material may be hidden from *in situ* analysis. For this reason, *ex situ* analysis of left-over residues is becoming increasingly popular (Nuevo et al. 2012, Materese et al. 2017, 2018). This has coincided with progress in analytical methodologies which have allowed greater sensitivities to be accessed, such as the advent of hyphenated techniques such as GC-MS/MS. *Ex situ* analysis does present some challenges due to the need to remove the sample from the astrochemical chamber, transport it to an analytical facility, prepare it for analysis (e.g., by derivatization), and all the while preclude unwanted physico-chemical changes to the residue (e.g., oxidation as a result of contact with ambient air). A properly validated analytical methodology should thus account for these factors (Fulvio et al. 2021).

Another factor which should be considered is the type of energetic processing available at individual research facilities. As discussed previously, astrochemical phenomena may be viewed as individual processes occurring on an event-by-event basis, making sequential processing studies using processing methods of different types a potentially interesting route for laboratory studies. However, if such an investigation is to be performed, or indeed, if the type of processing used in an experiment is to serve as one of the parameters in a statistical experimental design, then it is necessary for multiple energetic sources (e.g., ion beamlines, electron guns, ultraviolet lamps, etc.) to be available at the same research facility. The 'complete' astrochemistry experiment would thus have access to a range of processing types, including ion and electron sources with wide energy ranges alongside vacuum- and broadband ultraviolet photon sources. Other processing types, such as gamma and X-ray sources, could also be included.

This is, of course, an idealized situation, and it is highly unlikely that a single laboratory facility could host all such irradiation sources. However, some experimental astrochemistry groups have been successful in incorporating multiple energetic sources into a single set-up (e.g., Mifsud et al. 2021), while others have made their chambers portable, allowing for them to be transported to different facilities offering different processing types (e.g., Ioppolo et al. 2020). Such work-arounds may be the most cost-effective and technically feasible ways of incorporating multiple processing types into a statistical experimental design, thus allowing for their effects on the emergent properties of the astrochemical system to be studied and quantified.

## 4.3    Supervisory Control and Data Acquisition (SCADA)

For adequate control of a systems astrochemistry study, an appropriate control system and data architecture is required to overseen and manage the experiment. As an example, a supervisory control and data acquisition (SCADA) system architecture would allow for data acquisition from the whole experimental set-up, including peripherals, from a single human-machine interface. SCADAs are used extensively in manufacturing operations across many different industries and, if implemented, would open up exquisite control of the experimental rig with linked data acquisition, aggregation, and date/time stamping whilst also allowing open access to external collaborators. Consolidation of the collected data into a historian[3] would permit both real-time and remote access to the data for viewing and analysis. This is summarized in Fig. 10.

### 4.4   Data Analysis

The completed systems astrochemistry experiment will have generated a significant amount of data. As such, new approaches to analysis are required which will differ greatly to the univariate approach currently used by mainstream astrochemistry studies. The adaption of multi-variate techniques from the fields of chemometrics, machine learning, and observational astronomy may reveal insights into the data hitherto unmined. The use of a SCADA control system to manage the aggregation of data from multiple sensors would allow these data sets to be factorized using techniques such as singular value decomposition (SVD). For example, the FTIR data block (labelled as X) from an astrochemical experiment may be factorized into three matrices, as depicted in Fig. 11.

Matrix U contains information on the column space, while matrix V contains information on the row space and matrix $\Sigma$ is a diagonal matric matrix that provides information on the importance of U and V. SVD can therefore be used not only for data reduction but also to reveal hidden variables in the data (e.g., pseudo chemical composition). In a similar manner, multi-dimensional data blocks from multiple sensors and/or variables can be factorized using similar techniques. A data block might include factors such as spectroscopic or spectrometric data, acquisition time, temperature, and so on which then forms a multi-dimensional data cube that can be analyzed in a similar way to SVD, but using techniques such as higher order singular value decomposition (HOSVD). Machine learning algorithms could be used extensively with the resultant data sets.

Principal component analysis (PCA), which is essentially a truncated form of SVD, has gained some use in the field of observational astronomy where it has proven useful in revealing the relationship between the principal component and the ionization state (or lack thereof) of observed PAHs (Sidhu et al. 2020). A thorough description of all such possible techniques which may be employed in data analysis goes beyond the scope of this article, however, and the authors instead point the interested reader to the work of Brunton and Kutz (2019) for a deeper introduction to machine learning.

### 5   Modelling in the Context of Systems Astrochemistry: Introducing the Digital Twin

Conclusions within astrochemistry are often reached by linking together the results obtained by the three major research activities; laboratory experiments, observational astronomy, and astrochemical modelling. Such an approach has provided great insight into understanding the formation of COMs in the ISM. However, in putting forward the idea of a systems astrochemistry framework, we further propose that an additional consideration be made such that the experimentation occurring within laboratory astrochemistry chambers is also modelled via the use of a digital twin.

---

[3] The term 'historian', used extensively in industry, refers to the data repository for all operational and experimental data aggregation and storage.

A digital twin is a virtual representation of a physical system that serves as an *in silico* replica for scenario planning, sensitivity analysis, and modelling of responses to perturbations or changes in input and process parameters. The concept of a digital twin was first adopted in 2010 by NASA in an attempt to improve spacecraft simulation (Negri 2017) and has since been adopted widely within industrial settings, especially in manufacturing plants where digital twins support the digital transformation and adoption of Industry 4.0 concepts and model production efficiency.

Similarly, developing a digital twin for a laboratory astrochemistry experiment operating under a systems astrochemistry approach would accrue significant scientific benefits, particularly in the areas of sensitivity analysis and scenario planning. Thus, when underpinned by statistically constrained experimental designs, changes in input parameters and processing types could be successfully modelled. Examples of such modelling might include assessing the sensitivity of changes in energetic parameters to the resultant emergent properties or the *in silico* testing of experiments prior to experimental runs (which would be very useful to research groups with limited access to central research facilities). Additionally, insights into the chemistry related to the bounded nature of the experiment (e.g., astrochemical chamber wall effects) could also be gleaned.

## 6    Conclusions

Astrochemistry is a mature scientific discipline that is entering an exciting age with the deployment of both space-based (e.g., JWST) and ground-based (e.g., ALMA) telescope facilities capable of exploring the rich chemistry of the ISM. The integral role of chemistry in star and planet formation and the prevalence of prebiotic molecules in the ISM is becoming apparent and is revolutionizing our knowledge of the origins of life on Earth and the prospect of life being extant on exoplanets. Astrochemistry is therefore one of the most exciting fields of modern astronomy, if not science in general. The extension of astrochemistry to embrace ideas from planetary science and cosmochemistry would provide greater opportunities and would engage even larger communities.

Success in astrochemical research is a combination of observation, modelling, and laboratory studies each informing and challenging one another. In this article, we have reviewed the present status of laboratory astrochemistry and discussed a new approach based on systems chemistry requiring statistically designed experiments that consider all relevant experimental parameters (which we refer to in experimental design parlance as 'factors') simultaneously rather than the more traditional OFAT framework. This approach is needed to allow cross-correlation between different experiments and apparatus and thus avoid the 'more is less' facet of data production.

Thus, the systems chemistry approach focuses on emergent properties of the system, which in the context of systems astrochemistry is the final chemical products of the experiment. Systems astrochemistry acknowledges the complexity and multi-component nature of the chemistry, but allows for a more in-depth analysis of how the pathways towards such products depend on experimental conditions and constraints. That is to say, the physico-chemical constraints, role of ice morphology and temperature, grain structure, multiple methods of ice processing, and other factors should all be studied simultaneously in a structured systematic design to produce the conditions under which the formation of a molecular species occurs.

Adopting such a systems astrochemistry approach is expected to reveal insights into the physics and chemistry of the ISM beyond that garnered from current linear and sequential approaches. Furthermore, such a statistical approach is better matched with tests of astrochemical models allowing the integrity (completeness) and sensitivity analysis of such models to be tested against experiment, permitting digital twins of laboratory simulations to be developed.

In order to develop an experimental programme that is truly representative of a systems chemistry approach, it is necessary to carefully design the experiments with integrated equipment and control systems, robust hardware, and software architectures based on SCADA system architecture (or similar) that can analyse the data using multi-dimensional analytical techniques. As the number of data sets grows, the adoption of machine learning is likely to become increasingly useful, as is the method of chemometrics. Furthermore, the analytical technologies used to analyse complex systems and the automation systems used to adopt a high-throughput approach to systems astrochemistry are already in place and, if adopted, will undoubtedly make vital contributions to the discipline.

Through the adoption of these new methodologies, the astrochemistry community can realistically expect to answer some of the greatest remaining questions of modern astronomy and science including how (and where) the chemical ingredients needed for establishing living systems are synthesized and whether life is likely to be prevalent across the Universe.

## 7    Conflict of Interest

The authors declare that the research was conducted in the absence of any commercial or financial relationships that could be construed as a potential conflict of interest.

## 8    Author Contributions

All authors developed the concept of systems astrochemistry, and all authors were responsible for writing the manuscript.

## 9    Funding

The authors are grateful to have received funding from the Europlanet 2024 RI which has been funded by the European Union 2020 Research Innovation Programme under grant agreement No. 871149. DVM is the grateful recipient of a University of Kent Vice-Chancellor's Research Scholarship.

## 10    ORCID Identification Numbers

Nigel J. Mason:        0000-0002-4468-8324

Perry A. Hailey:        0000-0002-8121-9674

Duncan V. Mifsud:    0000-0002-0379-354X

## 11    References

Arumainayagam, C.R., Garrod, R.T., Boyer, M.C., Hay, A.K., Bao, S.T., Campbell, J.S., Wang, J., Nowak, C.M., Arumainayagam, M.R., Hodge, P.J. (2019). Extraterrestrial prebiotic molecules: photochemistry vs. radiation chemistry of interstellar ices. Chem. Soc. Rev. 48, 2293.

Asche, S., Copper, G.J.T., Keenan, G., Mathis, C., Cronin, L. (2021). A robotic prebiotic chemist probes long term reactions of complexifying mixtures. Nature Commun. 12, 3547.

Baratta, G.A., Leto, G., Palumbo, M.E. (2002). A comparison of ion irradiation and UV photolysis of $CH_4$ and $CH_3OH$. Astron. Astrophys. 384, 343.

Brunton, S.L., Kutz, J.N. (2019). Data-Driven Science and Engineering – Machine Learning, Dynamical Systems and Control. Cambridge University Press.

Burton, A.S., Stern, J.C., Elsila, J.E., Glavin, D.P., Dworkin, J.P. (2012). Understanding prebiotic chemistry through the analysis of extraterrestrial amino acids and nucleobases in meteorites. Chem. Soc. Rev. 41, 5459.

Cami, J., Bernard-Salas, J., Peeters, E., Malek, S.E. (2010). Detection of $C_{60}$ and $C_{70}$ in a young planetary nebula. Science 329, 1180.

Carota, E., Botta, G., Rotelli, L., Di Mauro, E., Saladino, R. (2015). Current advances in prebiotic chemistry under space conditions. Curr. Org. Chem. 19, 1963.

Cataldo, F., Pontier-Johnson, M.A. (2002). Recent discoveries in carbon black formation and morphology and their implications on the structure of interstellar carbon dust. Full. Nanotubes Carb. Nanostructures 10, 1.

Chuang, K.J., Fedoseev, G., Ioppolo, S., van Dishoeck, E.F., Linnartz, H. (2016). H-atom addition and abstraction reactions in mixed CO, $H_2CO$ and $CH_3OH$ ices – an extended view on complex organic molecule formation. Mon. Not. R. Astron. Soc. 455, 1702.

de Marcellus, P., Meinert, C., Nuevo, M., Filippi, J.J., Danger, G., Deboffle, D., Nahon, L., Le Sergeant d'Hendecourt, L., Meierhenrich, U.J. (2011). Non-racemic amino acid production by ultraviolet irradiation of achiral interstellar ice analogues with circularly polarized light. Astrophys. J. Lett. 727, L27.

Deamer, D., Dworkin, J.P., Sandford, S.A., Bernstein, M.P., Allamandola, L.J. (2002). The first cell membranes. Astrobiology 2, 371.

Deming, S.N., Morgan, S.L. (1993). Experimental Design: A Chemometric Approach. Elsevier Science.

Ding, J.J., Boduch, P., Domaracka, A., Guillous, S., Langlinay, T., Lv, X.Y., Palumbo, M.E., Rothard, H., Strazzulla, G. (2013). Implantation of multiply charged sulfur ions in water ice. Icarus 226, 860.

Dulieu, F., Congiu, E., Noble, J., Baouche, S., Chaabouni, H., Moudens, A., Minissale, M., Cazaux, S. (2013). How micron-sized dust particles determine the chemistry of our universe. Sci. Rep. 3, 1338.

Fulvio, D., Potapov, A., He, J., Henning, T. (2021). Astrochemical pathways to complex organic and prebiotic molecules: Experimental perspectives for in situ solid-state studies. Life 11, 568.

Garrod, R.T., Herbst, E. (2006). Formation of methyl formate and other organic species in the warm-up phase of hot molecular cores. Astron. Astrophys. 457, 927.

Garrod, R.T. (2013). Three-dimensional, off-lattice Monte Carlo kinetics simulations of interstellar grain chemistry and ice structure. Astrophys. J. 778, 158.

Gentili, P.L. (2020). Astrochemistry and the theory of complex systems. Proceedings of the Observatory for Astrochemical Kinetics and Related Aspects, Accademia delle Scienze, Rome (Italy).

Geppert, W.D., Larsson, M. (2013). Experimental investigations in to astrophysically relevant ionic reactions.

Gerin, M., Liszt, H., Neufeld, D., Godard, B., Sonnentrucker, P., Pety, J., Roueff, E. (2019). Molecular abundances in the diffuse ISM: $CF^+$, $HCO^+$, $HOC^+$, and $C_3H^+$. Astron. Astrophys. 622, 26.

Henderson, B.L., Gudipati, M.S. (2015). Direct detection of complex organic products in ultraviolet (Ly$\alpha$) and electron-irradiated astrophysical and cometary ice analogs using two-step laser ablation and ionization mass spectrometry. Astrophys. J. 800, 66.

Holtom, P., Dawes, A., Mukerji, R., Davis, M., Webb, S.M., Hoffmann, S.V., Mason, N.J. (2006). VUV photoabsorption spectroscopy of sulfur dioxide ice. Phys. Chem. Chem. Phys. 8, 714.

Hornekær, L., Baurichter, A., Petrunin, V.V., Field, D., Luntz, A.C. (2003). Importance of surface morphology in interstellar $H_2$ Formation. Science 302, 1943.

Hornekær, L., Baurichter, A., Petrunin, V.V., Luntz, A.C., Kay, B.D., Al-Halabi, A. (2005). Influence of surface morphology on $D_2$ desorption kinetics from amorphous solid water. J. Chem. Phys. 122, 124701.

Ioppolo, S., Kaňuchová, Z., James, R.L., Dawes, A., Jones, N.C., Hoffmann, S.V., Mason, N.J., Strazzulla, G. (2020). Vacuum ultraviolet photoabsorption spectroscopy of space-related ices: 1 keV electron irradiation of nitrogen- and oxygen-rich ices. Astron. Astrophys. 641, 154.

Islam, S., Powner, M.W. (2017). Prebiotic systems chemistry: Complexity overcoming clutter. Chem. 2, 470.

Islam, S., Bučar, D.K., Powner, M.W. (2017). Prebiotic selection and assembly of proteinogenic amino acids and natural nucleotides from complex mixtures. Nature Chem. 9, 584.

James, R.L., Ioppolo, S., Hoffmann, S.V., Jones, N.C., Mason, N.J., Dawes, A. (2020). Systematic investigation of $CO_2:NH_3$ ice mixtures using mid-IR and VUV spectroscopy – part 1: thermal processing. RSC Adv. 10, 37515.

Jørgensen, J.K., Favre, C., Bisschop, S., Bourke, T., van Dishoeck, E.F., Schmalzl, M. (2012). Detection of the simplest sugar, glycolaldehyde, in a solar-type protostar with ALMA. Astrophys. J. Lett. 757, L4.

Kim, E.K., Martin, V., Krishnamurthy, R.J. (2017). Orotidine-containing RNA: Implications for the hierarchical selection (systems chemistry emergence) of RNA. Chem. Eur. J. 23, 12668.

Krishnamurthy, R.J. (2020). Systems chemistry in the chemical origins of life: The 18[th] camel paradigm. J. Systems Chem. 8, 40.

Larsson, M., McCall, B.J., Orel, A.E. (2008). The dissociative recombination of $H_3^+$ – A saga coming to an end? Chem. Phys. Lett. 462, 145.

Larsson, M., Geppert, W.D., Nyman, G. (2012). Ion chemistry in space. Rep. Prog. Phys. 75, 066901.

Li, J., Nowak, P., Otto, S. (2013). Dynamic combinatorial libraries: From exploring molecular recognition to systems chemistry. J. Am. Chem. Soc. 135, 9222.

Linnartz, H., Ioppolo, S., Fedoseev, G. (2015). Atom addition reactions in interstellar ice analogues. Int. Rev. Phys. Chem. 34, 205.

Lo, J.I., Chou, S.L., Peng, Y.C., Lin, M.Y., Lu, H.C., Cheng, B.M. (2014). Photochemistry of solid interstellar molecular samples exposed to vacuum-ultraviolet synchrotron radiation. J. Electron. Spectrosc. Relat. Phenomena 196, 173.

Loeffler, M.J., Hudson, R.L. (2010). Thermally-induced chemistry and the Jovian icy satellites: A laboratory study of the formation of sulfur oxyanions. *Geophys. Res. Lett.* 37, L19201.

Loeffler, M.J., Hudson, R.L. (2016). What is eating ozone? Thermal reactions between $SO_2$ and $O_3$: Implications for icy environments. Astrophys. J. Lett. 833, L9.

Ludlow, R.F., Otto, S. (2008). Systems chemistry. Chem. Soc. Rev. 37, 101.

Mason, N.J., Dawes, A., Holtom, P.D., Mukerji, R.J., Davis, M.P., Sivaraman, B., Kaiser, R.I., Hoffmann, S.V., Shaw, D.A. (2006). VUV spectroscopy and photo-processing of astrochemical ices: An experimental study. Faraday Discuss. 133, 311.

Mason, N.J., Drage, E.A., Webb, S.M., Dawes, A., McPheat, R., Hayes, G. (2008). The spectroscopy and chemical dynamics of microparticles explored using an ultrasonic trap. Faraday Discuss. 137, 367.

Mason, N.J., Nair, B., Jheeta, S., Szymańska, E. (2014). Electron induced chemistry: A new frontier in astrochemistry. Faraday Discuss. 168, 235.

Materese, C.K., Nuevo, M., Sandford, S.A. (2017). The formation of nucleobases from the ultraviolet photo-irradiation of purine in simple astrophysical ice analogs. Astrobiology 17, 761.

Materese, C.K., Nuevo, M., McDowell, B.L., Buffo, C.E., Sandford, S.A. (2018). The photochemistry of purine in ice analogs relevant to dense interstellar clouds. Astrophys. J. 864, 44.

Mattia, E., Otto, S. (2015). Supramolecular systems chemistry. Nature Nanotech. 10, 111.

McGuire, B.A., Carroll, P.B., Loomis, R.A., Finneran, I.A., Jewell, P.R., Remijan, A.J., Blake, G.A. (2016). Discovery of the interstellar chiral molecule propylene oxide ($CH_3CHCH_2O$). Science 352, 1449.

McKellar, A. (1940). Evidence for the molecular origin of some hitherto unidentified interstellar lines. Publ. Astron. Soc. Pac. 52, 187.

Mejía, C.F., de Barros, A.L.F., Bordalo, V., da Silveira, E.F., Boduch, P., Domaracka, A., Rothard, H. (2013). Cosmic ray-ice interaction studied by radiolysis of 15 K methane ice with MeV O, Fe and Zn ions. Mon. Not. R. Astron. Soc. 433, 2368.

Mifsud, D.V., Juhász, Z., Herczku, P., Kovács, S.T.S., Ioppolo, S., Kaňuchová, Z., Czentye, M., Hailey, P.A., Traspas Muiña, A., Mason, N.J., McCullough, R.W., Paripás, B., Sulik, B. (2021). Electron irradiation and thermal chemistry studies of interstellar and planetary ice analogues at the ICA astrochemistry facility. Eur. Phys. J. D 75, 182.

Miller, S.L. (1953). Production of amino acids under possible primitive Earth conditions. Science 117, 528.

Miller, S.L. (1955). Production of some organic compounds under possible primitive Earth conditions. J. Am. Chem. Soc. 77, 2351.

Moore, M.H., Donn, B., Khanna, R., A'Hearn, M.F. (1983). Studies of proton-irradiated cometary-type ice mixtures. Icarus 54, 388.

Mullikin, E., van Mulbregt, P., Perea, J., Kasule, M., Huang, J., Buffo, C., Campbell, J., Gates, L., Cumberbatch, H.M., Peeler, Z., Schneider, H., Lukens, J., Bao, S.T., Tano-Menka, R., Baniya, S., Cui, K., Thompson, M., Hay, A., Widdup, L., Caldwell-Overdier, A., Huang, J., Boyer, M.C., Rajappan, M., Echebiri, G., Arumainayagam, C.R. (2018). Condensed-phase photochemistry in the absence of radiation chemistry. ACS Earth Space Chem. 2, 863.

Muñoz-Caro, G.M., Meierhenrich, U.J., Schutte, W.A., Barbier, B., Arcones Segovia, A., Rosenbauer, H., Thiemann, W.H.P., Brack, A., Greenberg, J.M. (2002). Amino acids from ultraviolet irradiation of interstellar ice analogues. Nature, 416, 403.

Muñoz-Caro, G.M., Dartois, E., Boduch, P., Rothard, H., Domaracka, A., Jiménez-Escobar, A. (2014). Comparison of UV and high-energy ion irradiation of methanol:ammonia ice. Astron. Astrophys. 556, 93.

Negri, E. (2017). A review of the roles of digital twin in CPS-based production systems. Procedia Manuf. 11, 939.

Nuevo, M., Milam, S.N., Sandford, S.A. (2012). Nucleobases and prebiotic molecules in organic residues produced from the ultraviolet photo-irradiation of pyrimidine in $NH_3$ and $H_2O+NH_3$ ices. Astrobiology 12, 295.

Öberg, K.I. (2016). Photochemistry and astrochemistry: Photochemical pathways to interstellar complex organic molecules. Chem. Rev. 116, 9631.

Piani, L., Tachibana, S., Hama, T., Tanaka, H., Endo, Y., Sugawara, I., Dessimoulie, L., Kimura, Y., Miyake, A., Matsuno, J., Tsuchiyama, A., Fujita, K., Nakatsubo, S., Fukushi, H., Mori, S., Chigai, T., Yurimoto, H., Kouchi, A. (2017). Evolution of morphological and physical properties of laboratory interstellar organic residues with ultraviolet irradiation. Astrophys. J. 837, 35.

Powner, M.W., Sutherland, J.D. (2011). Prebiotic chemistry: A new modus operandi. Phil. Trans. R. Soc. B 366, 2870.

Ross, J., Arkin, A.P. (2009). Complex systems: From chemistry to systems biology. Proc. Nat. Acad. Sci. USA 106, 6433.

Sandford, S.A., Nuevo, M., Bera, P.P., Lee, T.J. (2020). Prebiotic astrochemistry and the formation of molecules of astrobiological interest in interstellar clouds and protostellar disks. Chem. Rev. 120, 4616.

Siegenfeld, A.F., Bar-Yam, Y. (2020). An introduction to complex systems science and its applications. Complexity 6105872.

Sivaraman, B., Corey, C.S., Mason, N.J., Kaiser, R.I. (2007). Temperature-dependent formation of ozone in solid oxygen by 5 keV electron irradiation and implications for Solar System ices. Astrophys. J. 669, 1414.

Snyder, L.E., Buhl, D., Zuckerman, B., Palmer, P. (1969). Microwave detection of interstellar formaldehyde. Phys. Rev. Lett. 22, 679.

Stecher, T.P. (1965). Interstellar extinction in the ultraviolet. Astrophys. J. 142, 1683.

Stecher, T.P. (1969). Interstellar extinction in the ultraviolet. II. Astrophys. J. 157, L125.

van Dishoeck, E.F. (2014). Astrochemistry of dust, ice and gas: Introduction and overview. Faraday Discuss. 168, 9.

Wakelam, V., Bron, E., Cazaux, S., Dulieu, F., Gry, C., Guillard, P., Habart, E., Hornekær, L., Morisset, S., Nyman, G., Pirronello, V., Price, S.D., Valdivia, V., Vidali, G., Watanabe, N. (2017). $H_2$ formation on interstellar dust grains: The viewpoints of theory, experiments, models and observations. Mol. Astrophys. 9, 1.

Watanabe, N., Kimura, Y., Kouchi, A., Chigai, T., Hama, T., Pirronello, V. (2010). Direct measurements of hydrogen atom diffusion and the spin temperature of nascent $H_2$ molecule on amorphous solid water. Astrophys. J. Lett. 714, 233.

Weinreb, S. Barrett, A.H., Meeks, M.L., Henry, J.C. (1963). Radio observations of OH in the interstellar medium. Nature 200, 829.

Whitesides, G.M., Ismagilov, R.F. (1999). Complexity in chemistry. Science 284, 89.

Widicus Weaver, S.L. (2019). Millimeter wave and submillimeter wave laboratory spectroscopy in support of observational astronomy. Annu. Rev. Astron. Astrophys. 57, 79.

Yakshinsky, B.V., Madey, T.E. (2001). Electron- and photon-stimulated desorption of K from ice surfaces. J. Geophys. Res. Planets 106, 33303.

Yocum, K.M., Smith, H.H., Todd, E.W., Mora, L., Gerakines, P.A., Milam, S.L., Widicus Weaver, S.L. (2019). Millimeter/submillimeter spectroscopic detection of desorbed ices: A new technique in laboratory astrochemistry. J. Phys. Chem. A 123, 8702.

## 12    Figures and Figure Captions

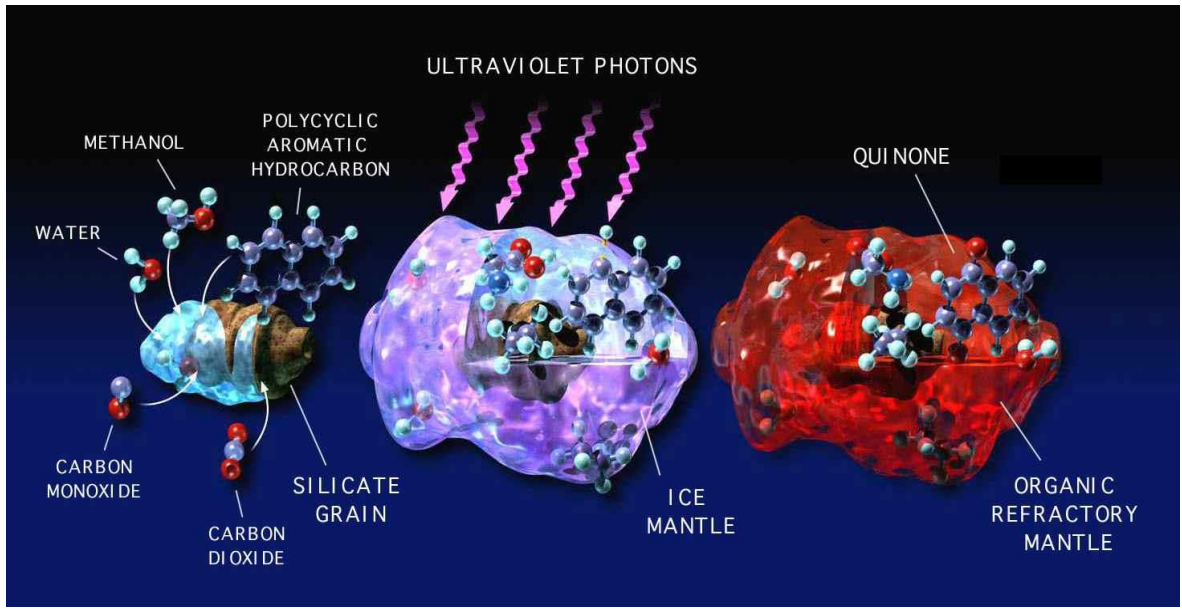

**Fig. 1** The Greenberg Model of interstellar ice mantle formation and chemical evolution. The ice mantle grows by condensation of gas-phase species onto the surfaces of cold dust grains. Simultaneously, surface reactions between these species, ultraviolet irradiation, and galactic cosmic ray bombardment allow for a complex chemistry in the ice mantle to develop which leads to the formation of the complex molecular species observed in the ISM. Figure reproduced with permission from Sandford et al. (2020).

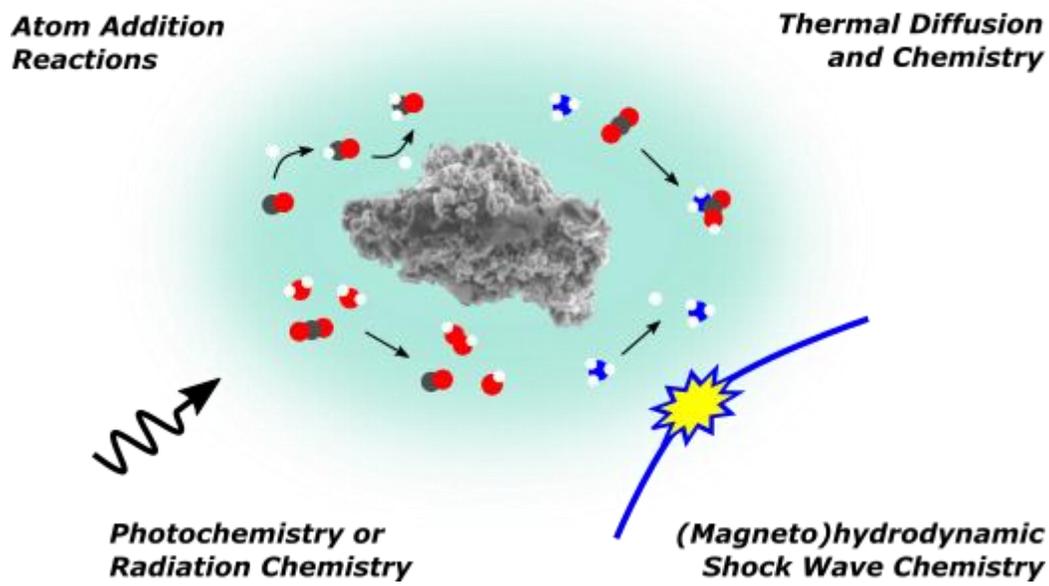

**Fig. 2** Five mechanisms by which complex molecular species may be synthesized in the ice mantles on interstellar dust grains. Thickness of ice mantle is exaggerated for visualization purposes.

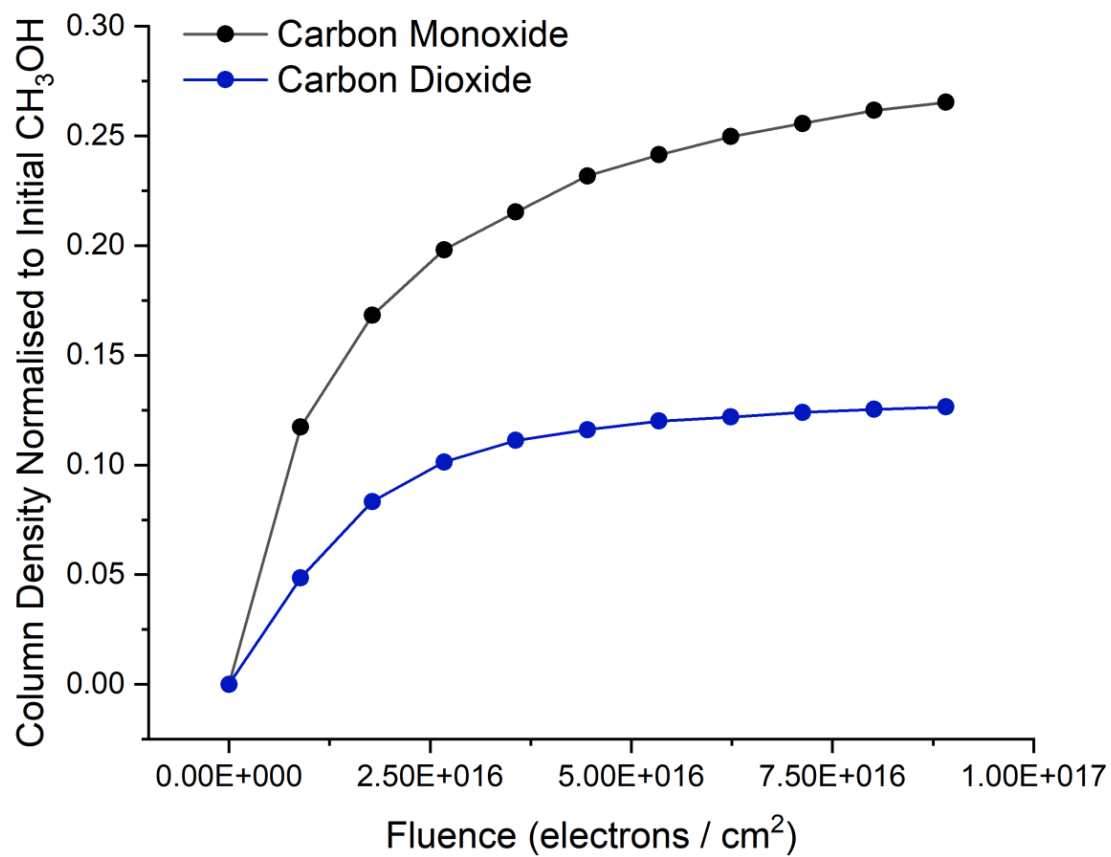

**Fig. 3** Formation of CO and CO$_2$ after electron irradiation of amorphous CH$_3$OH at 20 K. As can be seen, the abundances of the products increase significantly at first, only to begin to plateau at higher fluences.

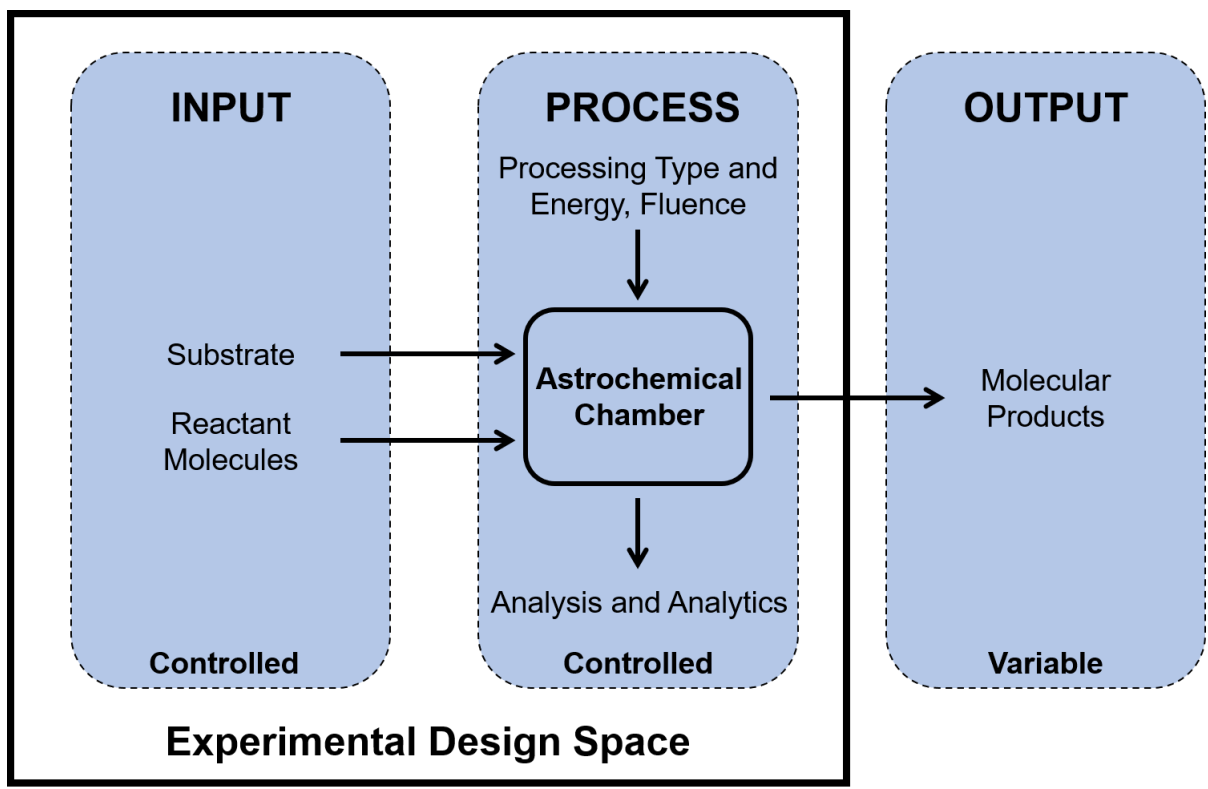

**Fig. 4** Example of an astrochemical design space concept visualized as an IPO sequence.

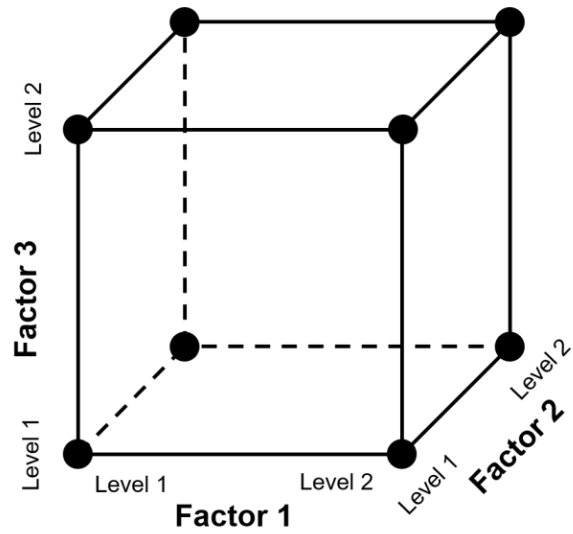

**Fig. 5** A simple example of a full factorial experimental design in which three factors are investigated at two levels. The resultant eight runs are represented by the corners of the experimental design space.

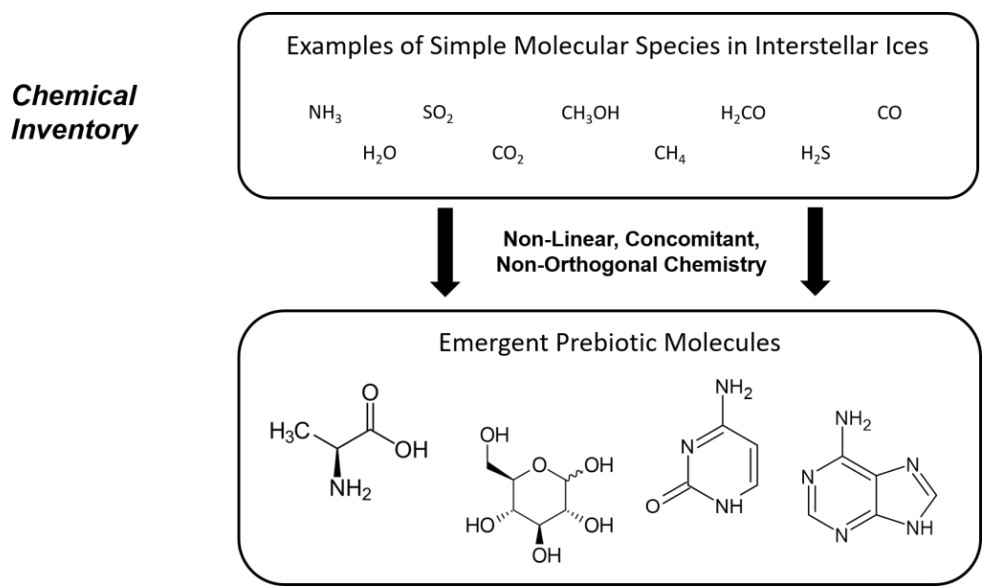

**Fig. 6** The evolution of the dense dust cloud and the associated change in conditions to which ice grain mantles are exposed mean that COM formation is not a linear or orthogonal process, likely involving several production and destruction phases.

# Electron and Photon Processing of Pure and Mixed Ices

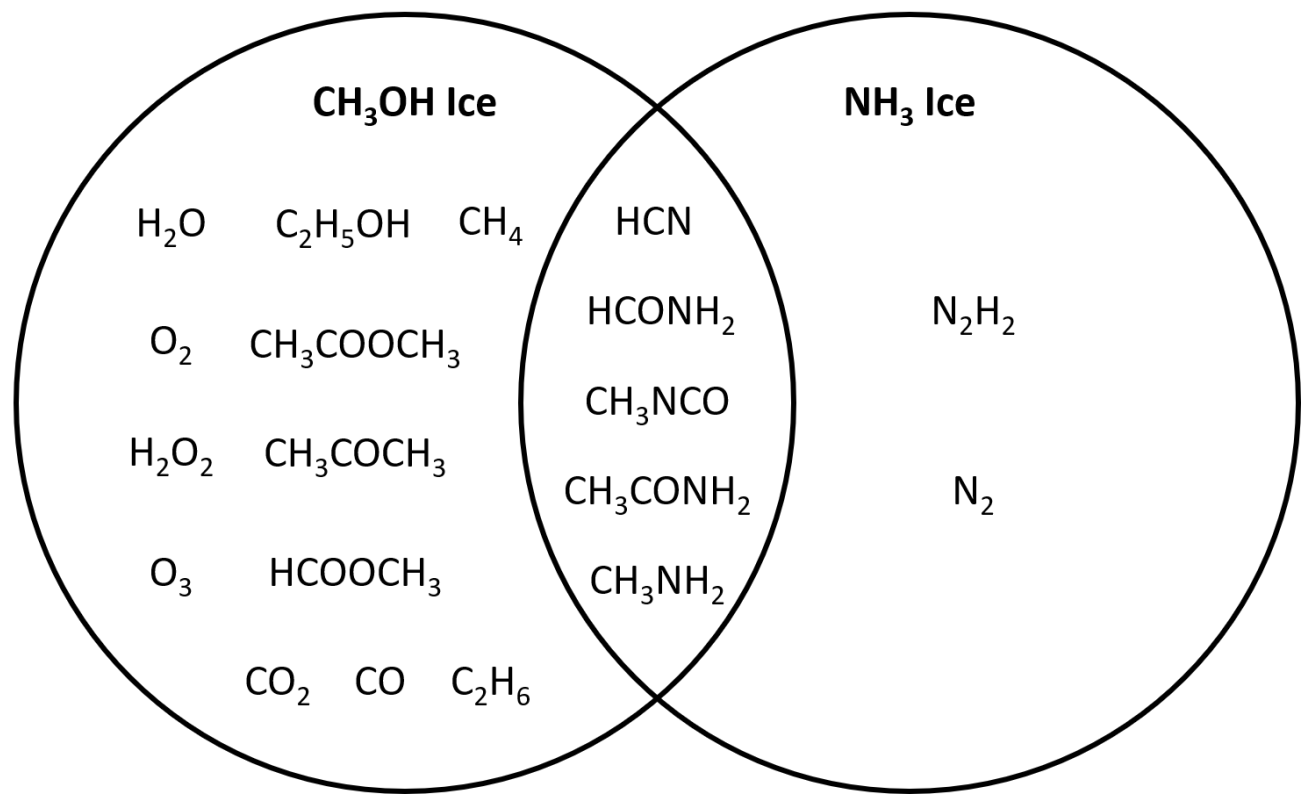

**Fig. 7** The irradiation of simple mixtures of molecules such as $H_2O$, $CH_3OH$, and $NH_3$ may result in a wealth of more complex molecules which, upon further processing, may themselves go on to form COMs of prebiotic interest. Data amalgamated from various sources.

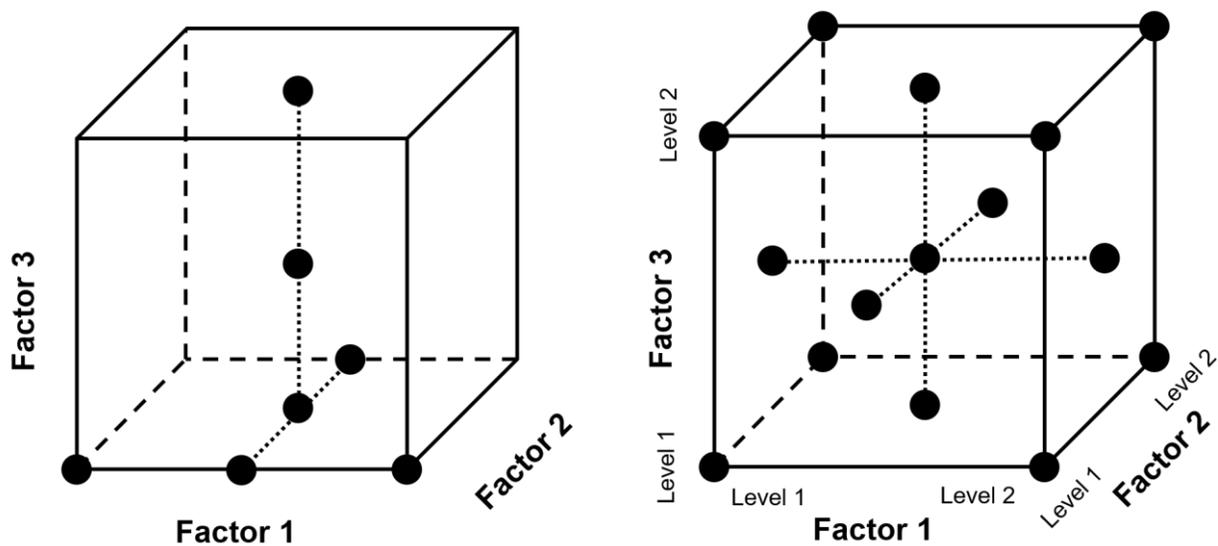

**Fig. 8** The astrochemical design space is less fully mapped in an OFAT study of three experimental parameters (left) compared to that of a full factorial experimental design at two levels with additional center point analysis replicates (right).

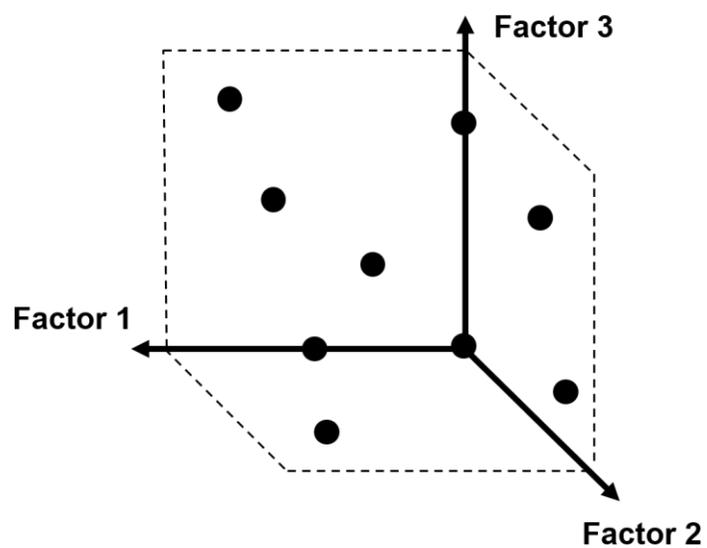

**Fig. 9** A three-dimensional scatter plot analyzing the responses documented during a statistical experimental design study looking into three factors.

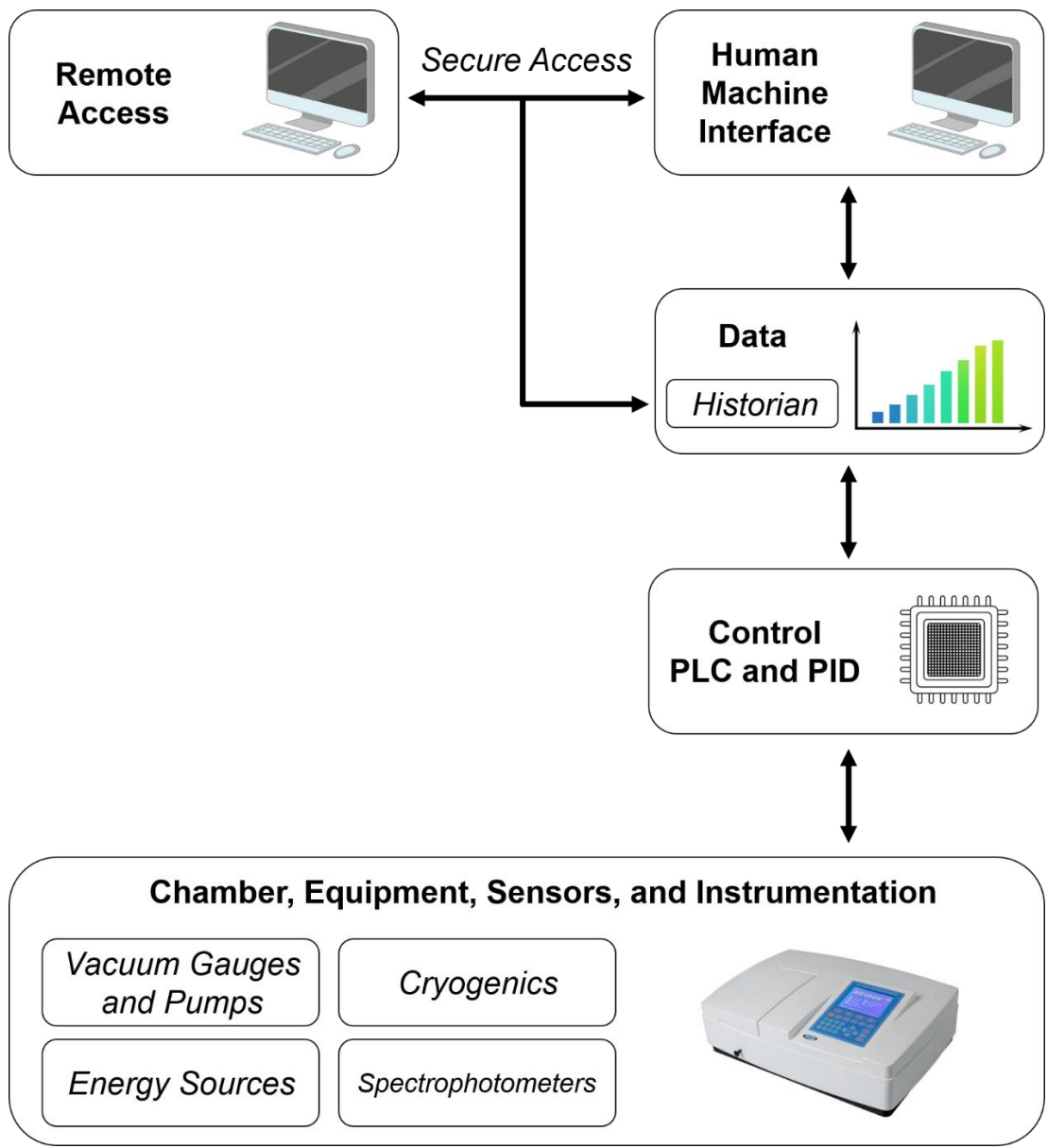

**Fig. 10** A representation of a SCADA system layout of the type proposed to be incorporated into systems astrochemistry experiments.

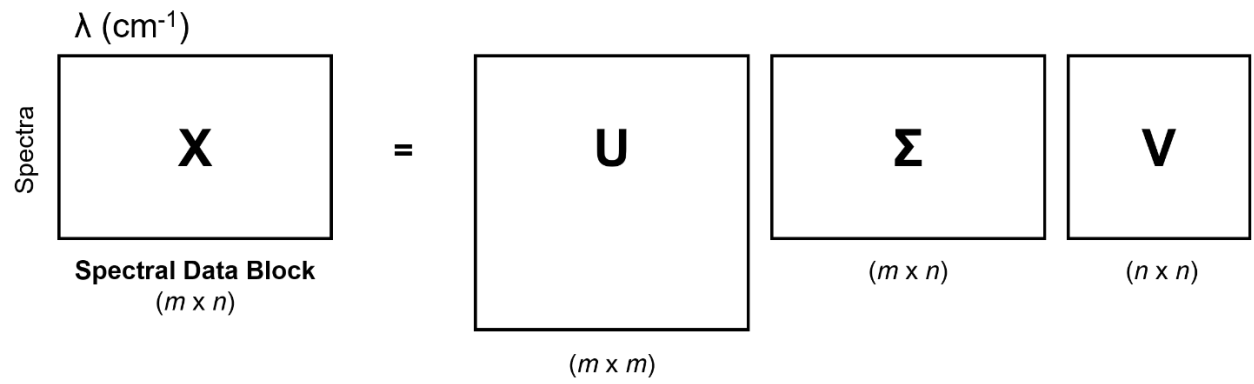

**Fig. 11** Graphical representation of singular value decomposition (SVD) of a spectral data block.

## 13    Tables

**Table 1** Example of the different parameters and levels used during an ice processing study making use of a full factorial experimental design.

| Factor | Unit | Levels | |
|---|---|---|---|
| | | **Low** | **High** |
| Ice Thickness | μm | 0.1 | 3.0 |
| Ice Morphology | N/A | Amorphous | Crystalline |
| Processing Energy | keV | 5 | 100 |
| Fluence Reached | $cm^{-2}$ | $10^{14}$ | $10^{16}$ |
| Temperature | K | 20 | 80 |

**Table 2** List of runs to be performed and the levels of each parameter investigated for the thought experiment in which a pure ice is processed with five experimental factors considered at two levels. All possible factor and level combinations are explored across the 32 runs, with a further three runs serving as center point analysis replicates. Run order is also randomized in an effort to preclude systematic errors.

| Exp No | Run Order | Ice Thickness | Morphology | Energy | Fluence | Temperature |
|---|---|---|---|---|---|---|
| 1 | 26 | 0.1 | Amorphous | 5 | $10^{14}$ | 20 |
| 2 | 35 | 3.0 | Amorphous | 5 | $10^{14}$ | 20 |
| 3 | 19 | 0.1 | Crystalline | 5 | $10^{14}$ | 20 |
| 4 | 3 | 3.0 | Crystalline | 5 | $10^{14}$ | 20 |
| 5 | 22 | 0.1 | Amorphous | 100 | $10^{14}$ | 20 |
| 6 | 23 | 3.0 | Amorphous | 100 | $10^{14}$ | 20 |
| 7 | 34 | 0.1 | Crystalline | 100 | $10^{14}$ | 20 |
| 8 | 8 | 3.0 | Crystalline | 100 | $10^{14}$ | 20 |
| 9 | 17 | 0.1 | Amorphous | 5 | $10^{16}$ | 20 |
| 10 | 16 | 3.0 | Amorphous | 5 | $10^{16}$ | 20 |
| 11 | 30 | 0.1 | Crystalline | 5 | $10^{16}$ | 20 |
| 12 | 5 | 3.0 | Crystalline | 5 | $10^{16}$ | 20 |
| 13 | 6 | 0.1 | Amorphous | 100 | $10^{16}$ | 20 |
| 14 | 25 | 3.0 | Amorphous | 100 | $10^{16}$ | 20 |
| 15 | 33 | 0.1 | Crystalline | 100 | $10^{16}$ | 20 |
| 16 | 1 | 3.0 | Crystalline | 100 | $10^{16}$ | 20 |
| 17 | 29 | 0.1 | Amorphous | 5 | $10^{14}$ | 80 |
| 18 | 12 | 3.0 | Amorphous | 5 | $10^{14}$ | 80 |
| 19 | 2 | 0.1 | Crystalline | 5 | $10^{14}$ | 80 |
| 20 | 13 | 3.0 | Crystalline | 5 | $10^{14}$ | 80 |
| 21 | 9 | 0.1 | Amorphous | 100 | $10^{14}$ | 80 |
| 22 | 10 | 3.0 | Amorphous | 100 | $10^{14}$ | 80 |
| 23 | 31 | 0.1 | Crystalline | 100 | $10^{14}$ | 80 |
| 24 | 20 | 3.0 | Crystalline | 100 | $10^{14}$ | 80 |
| 25 | 28 | 0.1 | Amorphous | 5 | $10^{16}$ | 80 |
| 26 | 15 | 3.0 | Amorphous | 5 | $10^{16}$ | 80 |
| 27 | 11 | 0.1 | Crystalline | 5 | $10^{16}$ | 80 |
| 28 | 24 | 3.0 | Crystalline | 5 | $10^{16}$ | 80 |
| 29 | 14 | 0.1 | Amorphous | 100 | $10^{16}$ | 80 |
| 30 | 32 | 3.0 | Amorphous | 100 | $10^{16}$ | 80 |
| 31 | 4 | 0.1 | Crystalline | 100 | $10^{16}$ | 80 |
| 32 | 7 | 3.0 | Crystalline | 100 | $10^{16}$ | 80 |
| 33 | 21 | 1.5 | Crystalline | 50 | $10^{15}$ | 50 |
| 34 | 27 | 1.5 | Crystalline | 50 | $10^{15}$ | 50 |
| 35 | 18 | 1.5 | Crystalline | 50 | $10^{15}$ | 50 |